\documentclass[11pt,a4paper]{article}

\usepackage{amsmath,amssymb,graphicx}
\usepackage{mcite}
\usepackage{color}

\newcommand{\ii}{\mathrm{i}}

\newcommand{\dd}{\mathrm{d}}

\newcommand{\e}{\mathrm{e}}

\newcommand{\tr}{\mathop{\mathrm{tr}}\nolimits}

\newcommand{\I}{\mathbb{I}}

\begin{document}
%

\title{Four-dimensional gauge and gravity models from texture graphs}%
\author{Corneliu Sochichiu
\thanks{E-mail:
corneliu@gist.ac.kr}\\
GIST College, Gwangju Institute of Science and Technology,\\
123 Cheomdan-gwagiro(Oryong-dong), Buk-gu\\ 
Gwangju 500-712 Republic of Korea
}%
\maketitle

\begin{abstract}
We study statistical graph models leading in continuum limit to relativistic fermionic fields coupled to gravity and gauge fields in four-dimensional space-time. 
\end{abstract}

\section{Introduction}\label{sec:Intro}
Quantum field theory models rely on three basic principles: Lorentz invariance, continuous space-time, and locality. Although  firmly established empirically, this combination of principles was in the past and still is a source of mathematical and conceptual difficulties. Most of mathematical issues, e.g. UV divergencies were successfully solved within the the quantum field theory framework. However, when it comes to gravity in which the local Lorentz symmetry extended to the diffeomorphism invariance assumes the role of gauge symmetry, there is no successful model combining all three principles. Moreover, there is an argument~\cite{Penrose:1972jq}, according to which a quantum gravity theory combining all these concepts is in principle impossible. Hence, a successful theory of gravity should possess these properties only in an approximate way which is precise only within the experimental window available to us today.

Nevertheless, there is string theory which successfully quantises gravity, albeit with no locality. In fact, the non-perturbative approach to strings opens a Pandora's box of models, all of which are unified under the paradigm of M-theory, which itself is not yet a fully understood model. In particular, there are string theory inspired matrix models in which the space-time emerges as an arena for large scale dynamics (see Ref.~\cite{Kim:2011cr}, for some recent results).

Although mathematically elegant and feature rich, string theory fails to fit naturally with Standard Model or other experimentally based particle theories. The problem is that it predicts too many structures, while those compatible with observations are by no means special~\cite{Douglas:2003um}. What is also true, there is a wide scale gap between the regime claimed by string theory and our present day experimental knowledge as it is described by QFT models.

On the other hand, in condensed matter theory we have situations that are well understood from both macroscopic and microscopic aspects. A condensed matter system is characterised by a hierarchy of structures where at the `microscopic' level we have atomic structures resulting from `more fundamental' electronic exchange forces organised at lower energies into `less fundamental' collective excitation modes naturally described in field theory frameworks. 
Particular interest present so called Dirac materials, like graphene (see e.g.~\cite{2009RvMP...81..109C} for a review), in which Lorentz invariant models emerge in the macroscopic description.

Under deeper mathematical scrutiny the emergence of Lorentz invariant Dirac, Weyl or Majorana fermions in such models does not seem accidental. In fact, it is known for some time, that whenever the Fermi surface shrinks to a point, low energy fluctuations organise themselves into relativistic spinor objects~\cite{PhysRevLett.95.016405,volovik2009universe}. This observation is based on  ABS construction~\cite{Atiyah19643}.

In this work we assume that a quantum field theory model emerges as macroscopic effective theory from a microscopic structure, or texture, which we can describe as a graph. We don't question the origin of this structure, but try to establish the properties of the graph which would lead to relativistic fermionic model in continuum limit.

The difference of our model from traditional discretisation techniques is that in our case the entire space-time emerges from a statistical model on such a graph. Our observation is that the space-time containing a Dirac fermion emerges naturally as soon as we require linear non-degeneracy of fermionic fluctuations around saddle points in an analytically continued gaussian statistical model related to a translational invariant graph. We dub these points pseudo-Fermi points (pFp). The continuum limit is determined by the spectrum of the model near these points. We show that there is a choice of parameters in which we get a single species of Dirac fermion with global U(1) symmetry. Furthermore, continuous deformations of the graph within the class of translational invariant graphs lead to transitions modifying the fermionic spectrum and, respectively, the global symmetry group to non-abelian groups. On the other hand, deformations which break the translational invariance lead in the continuum limit to coupling of the fermion to external gravity and gauge fields. Surprisingly, in the case of non-symmetric placement of pFp in the Brillouin zone a chiral gauge coupling emerges as well. This is in strong contrast with the standard lattice discretisation approach.

The plan of the remaining of this paper is as follows. In the next section we set up the model and specify the type of graphs which we study. In the third section we study the occurrence and properties of pseudo-Fermi points. In the section 4. we describe graph deformations breaking the translational symmetry and obtain coupling to external gauge and gravity fields.

\section{The model}
As mentioned in the Introduction, there are graph models, called Dirac lattices, in which the low energy effective theory possesses Lorentz invariance~\cite{Sochichiu:2011ap}. The time extension in these models, however, is inherited from the absolute non-relativistic time of the microscopic observer. Deformations of such graphs lead to coupling of the fermion field to external gauge and gravity fields~\cite{Sochichiu:2010ns}  (see also \cite{Vozmediano2010}), but the background geometry can not be generic due to the absolute nature of time. Therefore, it is justified to attempt a generalisation of the approach to models where the entire space-time emerges, not only its spacial part.

The problem, which arises is that as soon as we have no time extension in the model, there is also no notion of energy, and as a sequence no concept of Fermi surface, and Fermi surface plays a fundamental role in models with relativistic continuum limit. 

To circumvent this difficulty, we have to define the partition function as a sum of complex phase factors rather then a sum of exponentially suppressed factors. So, consider the statistical model on a graph described by its Hermitian adjacency matrix \( T \), and partition function,
\begin{equation}\label{eq:PartitionFunction}
 	Z(T)= \int [\dd \psi^{\dag}][\dd \psi]\e^{\ii S[ \psi^{\dag}, \psi;T]},
\end{equation}
where \( [\dd \psi]=\prod_{k\in\text{Graph nodes}}\dd \psi_{k} \)  is the fermionic Grassmann measure and, correspondingly, \( [\dd \psi^{\dag}] \) its conjugate. The \emph{action} is given in terms of \( T \),
\begin{equation}\label{eq:Action}
 	S[ \psi^{\dag}, \psi;T]=
	\psi^{\dag} \cdot T \cdot \psi.
\end{equation} 
Such a sum is not properly defined and requires an analytic continuation for a complete definition. The analytic continuation is discussed below. It will induce a hierarchy of states relevant for the continuum limit and will allow the definition of a structure similar to fermi liquid.

The matrix \( T \) is the main focus of our study. We assume that the above graph structure is in broad equilibrium (as dictated by the microscopic theory). Therefore, the main contribution to the graph's dynamics, when we include it into consideration, should come from the fermionic back reaction. We consider the situation in which exactly one-half of states are excited. This is an analogue of half filling in condensed matter systems. 

Furthermore, we assume that statistically relevant contributions come from graph configurations leading to smooth local geometries and renormalisable interactions in the saddle point approximation. This belief is based on the role of RG flow~\cite{Wilson:1971bg}, according to which the deformations moving away from local and continuous quantum field theory are given by operators which are irrelevant in the continuum limit. Throughout this work we use this `quantum Darwinistic principle' as a filter to keep only relevant contribution. 

We start with graphs of a special structure and study the conditions under which they lead to relativistic fermion models in the continuum limit. The graphs we consider are (i) invariant under a \( D \)-dimensional translation group, i.e. split in blocks (unit cells) forming a \( D \)-dimensional lattice, and (ii) local, i.e. they have nontrivial adjacency only between nearest neighbour cells. Translational invariance is convenient for the mathematical description of the continuum limit, but as soon as a map between discrete and continuous quantities is established it can be broken by local graph deformations. 

All in all, the adjacency matrix for a translational invariant local graph takes the form,
\begin{equation}
 	T= \alpha + \sum_{I}( \alpha^{I} T^{I}+ \alpha^{\dag I} T_{I}^{\dag}),
\end{equation}
where \( T_{I} \), \( I=1,\dots D \), are the generators of translations while \( \alpha \), \( \alpha^{I} \), and \( \alpha^{\dag I} \) are, respectively, the Hermitian unit cell internal adjacency and the adjacency to the neighbour cell in the direction \( I \). 

Here we consider the smallest non-trivial unit cell. The case of one site unit cell is degenerate for any \( D>1 \).  Therefore, the simplest non-trivial case is that of unit cell consisting of two sites. In this case \( \alpha \) and \( \alpha^{I} \) are \( 2 \times 2 \) matrices. Different cases will be considered elsewhere.

In what follows, we consider the saddle point limit of such graphs and show that within some range of parameters  in the large scale limit these graphs generically lead to a free relativistic fermionic model. Then we show that a deformation of a such translational invariant graph by an arbitrary local operator having smooth limit leads to the coupling of the Dirac fermion to background geometry and gauge field. More technical details behind the results presented here will be reported in a separate manuscript~\cite{Sochichiu:inprogress}.

\section{Pseudo-Fermi points and Dirac fermion}\label{sec:sec}
 Let us consider the saddle point limit of the partition function~\eqref{eq:PartitionFunction}. Since the action is quadratic in the fermionic variable, the main contribution in this limit comes from configurations close to zero modes of the matrix \( T \). In the Fourier basis given by functions,
\begin{equation}
 	\psi(k)=\sum_{ \mathbf{n}} 
	\e^{\ii k \cdot \mathbf{n}} \psi_{ \mathbf{n}},
\end{equation}
where \( -\pi\leq k_{I} <\pi \) is the first Brillouin zone,  the adjacency matrix becomes a \( 2 \times 2 \) matrix function. It can be expanded in terms of the extended set of Pauli matrices \( \sigma^{A} \): \( T(k)=T_{A}(k) \sigma^{A} \),  (\( A=0,1,2,3 \)), where
\begin{equation}\label{adj-matr-transl-inv}
 	T_{A}(k)= \alpha_{A}+\sum_{I}
	\left(
	\alpha^{I}_{A}\e^{-\ii k_{I}}+ \bar{ \alpha}{}^{I}_{A}\e^{\ii k_{I}}
	\right).
\end{equation}
In this basis the zero mode condition takes the form, 
\begin{equation}\label{eq:SaddlePointCondition}
 	\det T(k) \equiv T_{0}^{2}(k)-T_{1}^{2}(k)-T_{2}^{2}(k)-T_{3}^{2}(k)=0.
\end{equation}

Eq.~\eqref{eq:SaddlePointCondition} points to a possible way for the analytic continuation of the integral~\eqref{eq:PartitionFunction}. Thus, an exponentially suppressing factor emerges if the diagonal part of \( T \) becomes imaginary,
\begin{equation}\label{eq:WickRotation}
 	\sigma_{0}=\I \mapsto \ii \sigma_{0}.
\end{equation}
This modification is somehow similar to a Wick rotation in field theory. 

With this analytic continuation in mind, the statistically dominant configuration in the saddle point limit is given by an analogue of Fermi sea, which we will call pseudo-Fermi sea. The analogue of the Fermi surface in this case is given by a variety of momentum space points satisfying \( T_{A}(k)=0 \) for \( A=1,\dots, D \). We consider the situation when such points are isolated and linearly non-degenerate. The ABS construction~\cite{Atiyah19643} implies that fermionic fluctuations in vicinity of such points are organised into Weyl spinors (see also  Refs.~\cite{PhysRevLett.95.016405,volovik2009universe,Sochichiu:2011ap}).

Concerning the occurrence  of such points that will be called pseudo-Fermi points (pFp), engineering an adjacency matrix with a pFp at a desired location \( K \) is very easy. It can be accomplished by picking some coefficients \( \alpha^{I}_{A} \), then the coefficients \( \alpha_{A} \) are unambiguously determined by the condition \( T_{A}(K)=0 \),
\begin{equation}\label{eq:solution-for-alpha}
 	\alpha_{A}=-\sum_{I} 
	\left(
	\alpha^{I}_{A}\e^{-\ii K_{I}}+ \bar{ \alpha}{}^{I}_{A}\e^{\ii K_{I}}
	\right).
\end{equation}
The only condition restricting the choice of \(  \alpha^{I}_{A} \) and \( K_{I} \) is coming from the linear non-degeneracy. It requires the matrix,
\begin{equation}\label{def:dT-AI}
 	\frac{ \partial T_{A}}{ \partial k_{I}}=-\ii 
	\left(
	\alpha^{I}_{A}\e^{-\ii k_{I}}- \bar{ \alpha}{}^{I}_{A}\e^{\ii k_{I}}
	\right),
\end{equation}
to have the rank \( \geq D \) at \( k=K \). This implies that \( D \) can not exceed four. In the case of \( D=4 \), the non-degeneracy condition requests non-vanishing of the determinant of \( \partial T/ \partial k \)¥. Moreover, in this case any small change to parameters \( \alpha_{A} \) and \( \alpha^{I}_{A} \) results at most in a motion of the pFp without destroying it. Therefore, the pFp will exist for a continuous range of parameters. This is a manifestation of stability, resulting from the appropriate choice of the translational symmetry dictated by the ABS construction~\cite{Atiyah19643}. Hence, from now on we assume \( D=4 \). 

It might be asked whether the above point \( K \) is a unique pFp or if it can be made such by a clever choice of parameters. The answer in both cases is `No', and for that there is an argument similar to one used in the Nielsen-Ninomiya theorem~\cite{Nielsen:1980rz}. 

For a pFp \( K_{ \sigma} \), where \( \sigma=1,2,\dots,N \) is a label counting pFps, consider the following topological charge~\cite{PhysRevLett.95.016405,volovik2009universe},
\begin{equation}\label{top-charge}
 	n_{ \sigma}= \frac{1}{24 \pi^{2} }\int_{S^{3}_{ \sigma}}
	\tr (\dd TT^{-1}\wedge \dd TT^{-1}\wedge \dd TT^{-1}),
\end{equation}
where the integration is performed over a small momentum space sphere \( S^{3}_{ \sigma} \) surrounding the point \( K_{ \sigma} \). Taken the topological nature and its discrete value, the value of this charge can not be modified by small continuos changes of parameters of the graph. As we show below, the fluctuations around such a pFp are described by a Weyl fermion with chirality \( n_{ \sigma}=\pm 1 \). The charge \( n_{ \sigma} \) is restricted to \( \pm 1 \) due to the linear nature of pFp. For a more general map, \(  n_{ \sigma}\in \mathbb{Z} \), we expect that one ends up with a multiplet of Weyl fermions, but this would correspond to non-linear points.  This is an interesting topic to study, however, beyond the scope of this paper.

The compactness of the Brillouin zone implies that the sum over all points of the zone vanishes,
\begin{equation}\label{eq:0TotalCharge}
 	\sum_{ \sigma} n_{ \sigma}=0.
\end{equation}
The sign of the topological charge is determined by that of \( \det (\partial T/ \partial k) \) at \( K_{ \sigma} \). 
Therefore, in addition to the original pFp, there must be at least one more separate pFp with the opposite sign of \( \det (\partial T/ \partial k) \). 

Alternatively, a pFp can be regarded as a common point of intersection between four three-dimensional surfaces \( T_{A}(k)=0 \)  in \( \mathbb{R}^{4} \). We are unaware of any theory describing how we can control such intersections.
However, by using \emph{Mathematica} we were able to model a number of such intersections. Thus, the simplest case, apart from no intersection, is the two-point intersection with \( n_{1}=-n_{2} \). In particular, within the parameter range where the surfaces are small enough, their shapes approach that of spheres, and four three-spheres intersect by either zero or two points, one-point intersection being the limiting degenerate case.

Now we shall show that in the case of two point intersection the continuum effective action is given by a free Dirac fermion model. Since the main contribution to the partition function in this approximation comes from modes near pFps, we can replace the adjacency operator \( T(k) \) by its linear expansion near these points. Therefore, the action takes the form,
\begin{equation}\label{low-energy-action}
 	S=\sum_{ \sigma} \int \frac{\dd^{4} k}{(2\pi)^{D}} \psi_{ \sigma}^{\dag}  \left(\frac{ \partial T_{A}}{ \partial k_{I}}
	\right)_{k=K_{ \sigma}}k_{I} \sigma^{A} \psi_{ \sigma}+\dots,
\end{equation}
where we introduced the notation: \( \psi_{ \sigma}= \psi(K_{ \sigma}+k) \), and dots denote higher orders in \( k \).

At every pFp, we \emph{could} relate the local momentum \( k_{I} \) to the macroscopic Cartesian momentum \( q_{A} \) by a linear transformation with the matrix \( (\partial T/ \partial k)_{K_{ \sigma}} \). When the determinant of this matrix is positive, this is a legitimate transformation. However, when it is negative, and this happens for one of pFps, the orientation is not preserved. Nevertheless, even for this point we can correct it by a modification of the sign of an odd number of columns or rows. For example, the Cartesian momentum \( q_{A} \) can be related to the microscopic momentum \( k_{I} \) by,
\begin{equation}
 	q_{A}= h_{A}^{ \sigma I}k_{I},
\end{equation}
where the positive determinant matrix \( h^{ \sigma} \) is defined as follows,
\begin{equation}
 	h_{0}^{ \sigma I}=  \left(\frac{ \partial T_{0}}{ \partial k_{I}}\right)_{K_{ \sigma}}, \qquad
	h_{a}^{ \sigma I}= n_{ \sigma}\left(\frac{ \partial T_{a}}{ \partial k_{I}}\right)_{K_{ \sigma}}
\end{equation}
and \( a=1,2,3 \). A different choice would be related to ours by a rotation.

In terms of new variables the action~\eqref{low-energy-action} becomes,
\begin{equation}
 	S= \int \frac{\dd^{4} q}{(2\pi)^{D}} \Psi^{\dag} (\I \otimes \I q_{0}+ \sigma^{a} \otimes \sigma^{3} q_{a}) \Psi,
\end{equation}
where we introduced the four component Dirac spinor \( \Psi \)  given by,
\begin{equation}
 	\Psi_{ \sigma}=
 	(\det h_{ \sigma})^{-1/2}\psi_{ \sigma}.
\end{equation}
By identifying the Dirac matrix basis as,
\begin{equation}
 	\gamma^{0}= \I \otimes \sigma^{1}, \qquad 
	\gamma^{a}= \gamma^{0} \cdot (\sigma^{a} \otimes \sigma^{3})
	= -\ii  (\sigma^{a} \otimes \sigma^{2}),
\end{equation}
and by doing a continuous inverse Fourier transform, we can cast the effective action in the standard form of the free Dirac particle action,
\begin{equation}\label{low-dim-eff}
 	S_{\text{eff}}=
		-\ii  \int \dd^{4} x \bar{ \Psi}\gamma^{A} \partial_{A} \Psi,
\end{equation}
where the bar denotes the Dirac conjugate: \( \bar{ \Psi}= \Psi^{ \dag} \gamma^{0} \).
\section{Graph deformations}
We have just shown that a translational invariant adjacency matrix brings in the continuum limit a free Dirac fermion. And a small variation of parameters that does not break the translational invariance can not change the spectrum of the theory. Large modification of parameters, however, can produce transitions to phases with a different fermionic field content. But now, let us discuss what happens if we allow deformations which do not respect translational invariance. 

For example, let us consider an operator \( D \), whose coefficients are cell dependent,
\begin{equation}\label{non-translational-matr}
 	D_{ \mathbf{n} ,\mathbf{m}}= \alpha_{ \mathbf{n}} \delta_{ \mathbf{n}, \mathbf{m}} 
	+ \sum_{I}( \alpha_{ \mathbf{n}}^{I} \delta_{ \mathbf{n}
	+ \hat{I}, \mathbf{m}}+ \alpha_{ \mathbf{n}}^{\dag I} \delta_{ \mathbf{n}- \hat{I}, \mathbf{m}}).
\end{equation}

Following our `quantum Darwinistic principle' we look for such a class of coefficients \(  \alpha_{ \mathbf{n}} \) and \( \alpha^{I}_{ \mathbf{n}} \) for which a local continuum limit exists. More technical details of the analysis and tools are discussed in the ref.~\cite{Sochichiu:inprogress}, here we just quote the results of the continuum limit.

Interestingly, in the case of two pFps, there are two distinct situations depending on whether the matrices \( h^{ 1,2} \)  are equal or not. Thus, the kernel of the continuum operator corresponding to \eqref{non-translational-matr}, takes the form,
\begin{multline}\label{non-transl-inv:operator}
 	D_{ \sigma \sigma'}(x,y)=
	\varphi_{ \sigma \sigma'}(x)\delta ( h^{ \sigma} \cdot x- h^{ \sigma'} \cdot y) \\
	+ \ii \xi_{ \sigma \sigma'}^{I}(x) \frac{ \partial}{ \partial y^{I}} \delta ( h^{ \sigma} \cdot x- h^{ \sigma'} \cdot y) \\
	- \ii\xi_{ \sigma \sigma'}^{I}(y) \frac{ \partial}{ \partial x^{I}} \delta ( h^{ \sigma} \cdot x- h^{ \sigma'} \cdot y).
\end{multline}
The contribution to components \( \xi_{ \sigma \sigma'} \) and \( \varphi_{ \sigma \sigma'} \) comes from the Fourier modes of 
\(
 	\alpha_{ \mathbf{n}} 
	+ \sum_{I}( \alpha_{ \mathbf{n}}^{I} \e^{-\ii K_{ \sigma I}}
	+ \alpha_{ \mathbf{n}}^{\dag I}\e^{\ii K_{ \sigma' I}} ),
\)
and \( \alpha^{I}_{ \mathbf{n}}\e^{-\ii K_{ \sigma}} \), near \( K_{ \sigma}-K_{ \sigma'} \). Therefore, the off-diagonal part is non-local as long as matrices \( h^{ \sigma} \) are not equal,  \( h_{ \sigma}\neq h_{ \sigma'} \) for \( ( \sigma\neq \sigma') \). In this case it should be dropped,  and the resulting local continuum action takes the form,
\begin{equation}\label{non-transl-inv:action}
 	S_{\text{eff}}=\\
	\int \dd^{4}x \left(-\ii\bar{ \Psi} \xi^{I}_{A} \gamma^{A} \partial_{I} \Psi
	+\bar{ \Psi} \chi_{A} \gamma^{5}\gamma^{A} \Psi
	+\bar{ \Psi} \nu_{A} \gamma^{A} \Psi\right),
\end{equation}
where the fields \( \xi, \chi \), and \( \nu \) are independent continuum fields built from \( \varphi \) and \( \xi \).

Although the action~\eqref{non-transl-inv:action} does not not appear explicitly diffeomorphism invariant, it can be made such by an appropriate change of fields~\cite{kleinert2008multivalued},
\begin{equation}\label{ntr-inv-nonsym:action-final}
 	S_{\text{eff}}=
		-\ii  \int \dd^{4} x \bar{ \Psi}\gamma^{A} \nabla_{A} \Psi,
\end{equation}
where \( \gamma^{I} \) are the co-moving Dirac matrices in the vierbein \( \xi_{A}^{I} \),
\begin{equation}
 	\gamma^{I}= \xi_{A}^{I} \gamma^{A}, \qquad
	\{ \gamma^{I}, \gamma^{J}\}= 2G^{IJ} \equiv
	2\eta^{AB}\xi_{A}^{I} \xi_{B}^{J},
\end{equation}
and \( \nabla_{A} \) is the spinor covariant derivative in this background,
\begin{equation}
 	\nabla_{I}= \partial_{I}- \frac{1}{4} \omega_{ABI} \gamma^{A}\gamma^{B}+ \ii \gamma^{5} A_{I}+V_{I},
\end{equation}
\( \omega_{ABI} \) being the corresponding spin connection. The fields \( A_{I} \) and \( V_{I} \) are, respectively, the external axial and vector Abelian gauge fields. By not receiving off-diagonal contribution in~\eqref{non-transl-inv:operator}, in the case of pFp asymmetry, the action can not get a mass term or any external coupling which may produce it, as it happens in \( (2+1) \)-dimensional case~\cite{Sochichiu:2010ns}. 

By contrast,  in the symmetric case the off-diagonal terms in Eq.~\eqref{non-transl-inv:operator} are local and, therefore, produce a Yukawa coupling term,
\begin{equation}
 	\Delta S_{\text{symm}}=
	\int \dd^{4}x \phi \bar{ \Psi} \Psi,
\end{equation}
where \( \phi \) is a scalar field born by modes near momenta \( K_{ \sigma}-K_{ \sigma'} \) of~\eqref{non-transl-inv:operator}. On the other hand, in this case the contribution leading to axial coupling, is not possible. Therefore, \( A_{I} \equiv 0 \). It is also worth noting that the symmetry of pFps is not stable: Unless it is enhanced dynamically, a generic graph deformation drives the system away from the symmetric point.

\section*{Discussion}
In this work we studied graph structures leading in continuum limit to relativistic fermionic particle. 
We have shown that the simplest configurations of fermionic graphs  lead in a natural way to Lorentz invariant gauge models in four-dimensional space-time as low energy theories. Deformations from translationally symmetric graphs result in couplings of fermionic fields to gauge and gravity background fields. 

If we accept the idea the space-time together with gauge and gravity interactions are a result of some microscopic structure, similar to ones in condensed matter theory, then the gauge and gravity dynamics should be determined by two main factors. First, there should be some microscopic forces about which we can not say much, perhaps. except that there is no \emph{a priori} reason to think that they have the same symmetry which we observe in the low energy theory.  If this is the case, they are completely irrelevant at the scale of energies we consider the continuum limit. The second factor is the fermionic back reaction. Apart from anomalies, it produces gauge/diffeomorphism invariant dynamical parts through what is known as Sakharov's mechanism~\cite{Sakharov:1967pk}. Unlike the fermionic part, which did not depend on the detailed parameters of the graph except the number and types of pFps, the couplings of induced gauge and gravity interactions are expected to depend on the locations of pFps as well as on the slopes of adjacency at these points. This is an interesting topic to address in a future research.

We restricted the adjacency to the nearest cell neighbour to simplify the analysis and make it more transparent. In fact, it is possible to add various non-local operators without changing the large scale behaviour, provided they vanish quickly enough with cell separation distance. 

Including deformations beyond the nearest neighbour range increases the diversity of interaction terms, giving more control over the large scale theory. It makes the adjacency matrix an almost generic function, subject only to topological restrictions. Thus, graphs leading to Standard Model field content are hopefully obtainable once we know the singularity structure of the adjacency in the momentum space studied in Ref.~\cite{volovik2009universe}.

The analysis of non-local graphs as well as the that of the dependence of the gauge couplings on the details of the graph structure is another possible direction for a future research. It would be interesting to establish the range of the parameter space corresponding to Standard Model and to look for a matrix model in which the configurations from this range dominate in some regime.

\subsubsection*{Acknowledgements.} 
I benefited from useful discussions with many colleagues. I want to thank Ellis Lee, for the hard effort to make the language of this manuscript seem English.

\bibliographystyle{unsrt}
\bibliography{TextureGraph}

\begin{thebibliography}{10}

\bibitem{Penrose:1972jq}
R.~Penrose.
\newblock {ON THE NATURE OF QUANTUM GEOMETRY}.
\newblock In J.~Klauder, editor, {\em Magic Without Magic}, pages 333--354, San
  Francisco, 1972. Freeman.

\bibitem{Kim:2011cr}
Sang-Woo Kim, Jun Nishimura, and Asato Tsuchiya.
\newblock {Expanding (3+1)-dimensional universe from a Lorentzian matrix model
  for superstring theory in (9+1)-dimensions}.
\newblock {\em Phys.Rev.Lett.}, 108:011601, 2012.

\bibitem{Douglas:2003um}
Michael~R. Douglas.
\newblock {The Statistics of string / M theory vacua}.
\newblock {\em JHEP}, 0305:046, 2003.

\bibitem{2009RvMP...81..109C}
A.~H. {Castro Neto}, F.~{Guinea}, N.~M.~R. {Peres}, K.~S. {Novoselov}, and
  A.~K. {Geim}.
\newblock {The electronic properties of graphene}.
\newblock {\em Reviews of Modern Physics}, 81:109--162, January 2009.

\bibitem{PhysRevLett.95.016405}
Petr Ho\v{r}{}ava.
\newblock {S}tability of {F}ermi {S}urfaces and ${K}$ {T}heory.
\newblock {\em Phys. Rev. Lett.}, 95:016405, Jun 2005.

\bibitem{volovik2009universe}
G.E. Volovik.
\newblock {\em {T}he {U}niverse in a {H}elium {D}roplet}.
\newblock The International Series of Monographs on Physics. Oxford University
  Press, 2009.

\bibitem{Atiyah19643}
M.F. Atiyah, R.~Bott, and A.~Shapiro.
\newblock {C}lifford modules.
\newblock {\em Topology}, 3, Supplement 1(0):3 -- 38, 1964.

\bibitem{Sochichiu:2011ap}
Corneliu Sochichiu.
\newblock Dirac lattice.
\newblock {\em Journal of Physics A: Mathematical and Theoretical},
  46(1):015002, 2013.

\bibitem{Sochichiu:2010ns}
Corneliu Sochichiu.
\newblock {On the defect induced gauge and Yukawa fields in graphene}.
\newblock {\em Int.J.Mod.Phys.}, B26:1250055, 2012.

\bibitem{Vozmediano2010}
M.~A.~H. Vozmediano, M.~I. Katsnelson, and F.~Guinea.
\newblock {G}auge fields in graphene.
\newblock {\em Physics Reports}, 496,:109, March 2010.

\bibitem{Wilson:1971bg}
Kenneth~G. Wilson.
\newblock {Renormalization group and critical phenomena. 1. Renormalization
  group and the Kadanoff scaling picture}.
\newblock {\em Phys.Rev.}, B4:3174--3183, Nov 1971.

\bibitem{Tomonaga:1950zz}
S.~Tomonaga.
\newblock {Remarks on Bloch's Method of Sound Waves applied to Many-Fermion
  Problems}.
\newblock {\em Prog.Theor.Phys.}, 5:544--569, 1950.

\bibitem{Luttinger:1963zz}
J.~M. Luttinger.
\newblock {An Exactly Soluble Model of a Many-Fermion System}.
\newblock {\em J. Math. Phys.}, 4:1154--1162, 1963.

\bibitem{Sochichiu:inprogress}
Corneliu Sochichiu.
\newblock On universality of lorentz symmetry in fermionic graphs.
\newblock {\em {I}n preparation}.

\bibitem{Nielsen:1980rz}
Holger~Bech Nielsen and M.~Ninomiya.
\newblock {Absence of Neutrinos on a Lattice. 1. Proof by Homotopy Theory}.
\newblock {\em Nucl.Phys.}, B185:20, 1981.

\bibitem{kleinert2008multivalued}
H.~Kleinert.
\newblock {\em Multivalued Fields in Condensed Matter, Electromagnetism, and
  Gravitation}.
\newblock World Scientific, 2008.

\bibitem{Sakharov:1967pk}
A.~D. Sakharov.
\newblock {V}acuum quantum fluctuations in curved space and the theory of
  gravitation.
\newblock {\em Sov. Phys. Dokl.}, 12:1040--1041, 1968.

\end{thebibliography}
	
\end{document}